\documentclass[twocolumn,showpacs,preprintnumbers,amsmath,amssymb,superscriptaddress]{revtex4}

\usepackage{graphicx}
\begin{document}
\bibliographystyle{prsty}
\title{Photoemission and x-ray absorption studies of valence states in (Ni,Zn,Fe,Ti)$_{3}$O$_{4}$ thin films exhibiting photo-induced magnetization}

\author{M.~Kobayashi}
\affiliation{Department of Physics, University of Tokyo, 
7-3-1 Hongo, Bunkyo-ku, Tokyo 113-0033, Japan}
\author{Y.~Ooki} 
\affiliation{Department of Physics, University of Tokyo, 
7-3-1 Hongo, Bunkyo-ku, Tokyo 113-0033, Japan}
\author{M.~Takizawa}
\affiliation{Department of Physics, University of Tokyo, 
7-3-1 Hongo, Bunkyo-ku, Tokyo 113-0033, Japan}
\author{G.~S.~Song}
\affiliation{Department of Physics, University of Tokyo, 
7-3-1 Hongo, Bunkyo-ku, Tokyo 113-0033, Japan}
\author{A.~Fujimori}
\affiliation{Department of Physics, University of Tokyo, 
7-3-1 Hongo, Bunkyo-ku, Tokyo 113-0033, Japan}
\author{Y.~Takeda}
\affiliation{Synchrotron Radiation Research Center, Japan Atomic Energy Research Institute, 
Mikazuki, Hyogo 679-5148, Japan}
\author{K.~Terai}
\affiliation{Synchrotron Radiation Research Center, Japan Atomic Energy Research Institute, 
Mikazuki, Hyogo 679-5148, Japan}
\author{T.~Okane}
\affiliation{Synchrotron Radiation Research Center, Japan Atomic Energy Research Institute, 
Mikazuki, Hyogo 679-5148, Japan}
\author{S.-I.~Fujimori}
\affiliation{Synchrotron Radiation Research Center, Japan Atomic Energy Research Institute, 
Mikazuki, Hyogo 679-5148, Japan}
\author{Y.~Saitoh}
\affiliation{Synchrotron Radiation Research Center, Japan Atomic Energy Research Institute, 
Mikazuki, Hyogo 679-5148, Japan}
\author{H. Yamagami}
\affiliation{Synchrotron Radiation Research Center, Japan Atomic 
Energy Research Institute, Mikazuki, 
Hyogo 679-5148, Japan}
\author{M.~Seki}
\affiliation{Department of Electronic Engineering, School of Engineering, University of Tokyo, 
7-3-1 Hongo, Bunkyo-ku, Tokyo 113-8656, Japan}
\author{T.~Kawai}
\affiliation{Institute of Scientific and Industrial Research, Osaka University, 
Ibaraki, Osaka 567-0047, Japan}
\author{H.~Tabata}
\affiliation{Department of Electronic Engineering, School of Engineering, University of Tokyo, 
7-3-1 Hongo, Bunkyo-ku, Tokyo 113-8656, Japan}
\affiliation{Department of Bioengineering, School of Engineering, University of Tokyo, 
7-3-1 Hongo, Bunkyo-ku, Tokyo 113-8656, Japan}
\date{\today}

\begin{abstract}
By means of photoemission and x-ray absorption spectroscopy, we have studied the electronic structure of (Ni,Zn,Fe,Ti)$_{3}$O$_{4}$ thin films, which exhibits a cluster glass behavior with a spin-freezing temperature $T_f$ of $\sim 230$ K and photo-induced magnetization (PIM) below $T_f$. 
The Ni and Zn ions were found to be in the divalent states. Most of the Fe and Ti ions in the thin films were trivalent (Fe$^{3+}$) and tetravalent (Ti$^{4+}$), respectively. While Ti doping did not affect the valence states of the Ni and Zn ions, a small amount of Fe$^{2+}$ ions increased with Ti concentration, consistent with the proposed charge-transfer mechanism of PIM. 
\end{abstract}


\maketitle
Spinel ferrite oxides are typical magnetic materials, and show various magnetic properties depending on the composition. The most advantageous feature of spinels is that various ions can be placed either at the tetrahedral or octahedral sites of the lattice, which allows us to control the magnetic properties in flexible ways. 
Recently, it has been reported that spinel ferrite films, (Mg,Ti,Fe)$_{3}$O$_{4}$ \cite{muraoka1}, (Co,Zn)Fe$_{2}$O$_{4}$ \cite{muraoka2} and (Ni,Zn,Fe,Ti)$_{3}$O$_{4}$ \cite{seki}, show cluster glass behavior near room temperature (RT) and exhibit photo-induced magnetization (PIM) below the spin-freezing temperature $T_f$. 
PIM is an attractive phenomenon that can be used for practical applications such as magneto-optical memories. 
In these ferrite films, nonmagnetic ions such as Ti$^{4+}$ and Zn$^{2+}$ are considered to disrupt the long-range ferromagnetic order and to induce magnetic clusters of various sizes. 
The spin-frozen state near RT is expected to result from the frustration and randomness of cluster-cluster interactions. 
Light irradiation causes the melting of the frozen spins, resulting in the increase of the magnetization. There are little change in the magnetization following light irradiation in the (Co,Zn)Fe$_{2}$O$_{4}$ films \cite{muraoka2} while the (Mg,Ti,Fe)$_{3}$O$_{4}$ films exhibit large PIM \cite{muraoka1}. 
In (Ni,Zn,Fe,Ti)$_{3}$O$_{4}$ (NZFT) thin films, PIM has been found much larger than the (Co,Zn)Fe$_{2}$O$_{4}$ and (Mg,Ti,Fe)$_{3}$O$_{4}$ films and is enhanced with the increase of Ti concentration, as shown by the excitation intensity dependence of the zero-field cool (ZFC) magnetization with light irradiation \cite{seki}. 
From the excitation energy dependence of the PIM in NZFT, Seki {\it et al}. \cite{seki} have proposed that the inter-valence charge transfer Ti$^{4+}$+Fe$^{2+}$ $\rightarrow$ Ti$^{3+}+$Fe$^{3+}$ contributes to the observed enhancement of the PIM. 
Therefore, to elucidate the mechanism of the PIM, further systematic investigations concerning the role of the Ti$^{4+}$ ions in the PIM are desired. 
In this work, we have performed photoemission (PES) and x-ray absorption spectroscopy measurements on NZFT thin films in order to investigate the valence states of the constituent ions. The Fe and Ti $3d$ partial density of states have also been extracted by resonant photoemission spectroscopy. The results support the scenario of the inter-valence charge transfer.

\begin{figure}[b!]
\begin{center}
\includegraphics[width=8.8cm]{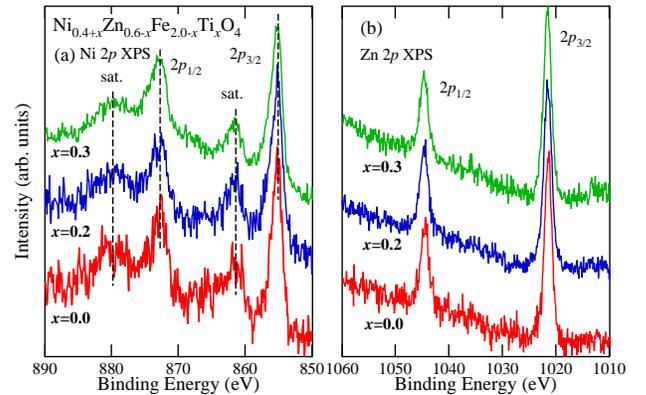}
\caption{Ni and Zn core-level XPS spectra of NZFT thin films with various Ti concentrations. 
(a) Ni $2p$ XPS spectra. Vertical lines are guide for the eyes. 
(b) Zn $2p$ XPS spectra. 
}
\label{Asite-XPS}
\end{center}
\end{figure}

A series of Ni$_{0.4+x}$Zn$_{0.6-x}$Fe$_{2-x}$Ti$_{x}$O$_4$ thin films with various Ti compositions ($x$=0.0, 0.2, and 0.3) were prepared on $\alpha$-Al$_2$O$_3$(0001) substrates by the pulsed-laser deposition (PLD) using an ArF excimer laser (wavelength: 193 nm). 
Here, the concentration of the non-magnetic ions (Zn + Ti) was kept constant in the series of samples in order to keep the cluster size nearly constant. 
The typical thickness of the films was 200 nm. The structural properties were examined with a standard x-ray diffraction (XRD) system using a monochromated Cu-$K\alpha$ radiation. 
Details of the sample preparation are described in Ref.~\cite{seki}. 
X-ray photoemission spectroscopy (XPS) measurements were performed using the Mg-$K\alpha$ line ($h\nu$ = 1253.6 eV). 
Resonant photoemission spectroscopy (RPES) and x-ray absorption spectroscopy (XAS) measurements were performed at the soft x-ray beamline BL23SU of SPring-8. XAS signals were detected by the total electron yield method. The XPS and RPES measurements were performed using a Gammadata Scienta SES-100 and SES-2000 hemispherical analyzer, respectively. 
The total energy resolution of XPS and RPES including temperature broadening was about $\sim 800$ and 200 meV, respectively. 
Binding energies were calibrated using the Au 4$f_{7/2}$ core-level peak at 84.0 eV. All the measurements were performed in the base pressure of  $\sim 1 \times 10^{-8}$ Pa at room temperature. 
The sample surfaces were cleaned by annealing at 400 $^\circ$C in oxygen atmosphere.

\begin{figure}[t!]
\begin{center}
\includegraphics[width=6.5cm]{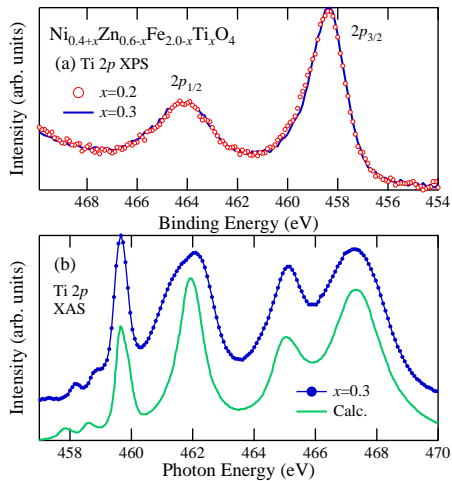}
\caption{Ti $2p$ core-level spectra of NZFT thin film. 
(a) Ti $2p$ XPS spectra. 
(b) Ti $2p$ XAS spectra. For comparison, the calculated spectrum has been reproduced from Ref.~\cite{abbate}. 
}
\label{Ti2p}
\end{center}
\end{figure}

Three kinds of ions are generally available for substitution in the regular spinel structure $AB_2$O$_4$: octahedral $A^{2+}$, octahedral $B^{3+}$, and tetrahedral $B^{3+}$ ions. In order to study Ti doping effects on the NZFT film, the Ti-concentration dependences of the valence states of the Ni and Zn ions were first examined. 
Figure~\ref{Asite-XPS} shows the Ni $2p$ and Zn $2p$ core-level XPS spectra. 
The Ni $2p$ spectra are independent of Ti concentration as shown in panel (a). The Ni $2p_{3/2}$ peak position is close to that of NiO \cite{geunseop, XPS}, indicating that the Ni ions are divalent. Also, the Zn $2p$ XPS spectra did not change with Ti doping as shown in panel (b). The binding energy of the Zn $2p_{3/2}$ peak is the typical value for Zn$^{2+}$ \cite{XPS}. The observations suggest that the valence states of the Ni and Zn ions substituting the $A$ site in NZFT are not affected by the Ti doping.

Figure~\ref{Ti2p} shows the Ti $2p$ XPS and XAS spectra of the NZFT thin films. 
The Ti $2p$ XPS spectrum of the $x$=0.2 sample is the same as that of the $x$=0.3 one as shown in panel (a), and they are characteristic spectra of the Ti$^{4+}$ state \cite{PRB_96_Morikawa}. 
The Ti $2p$ XAS spectrum is compared with that calculated for the Ti$^{4+}$ ion in an octahedral crystal field as in SrTiO$_{3}$ \cite{abbate}. The line shape of the Ti $2p$ XAS spectrum is similar to the calculated spectrum, indicating that the doped Ti ions are tetravalent and are located at the octahedral sites. 
The result is consistent with the conclusion from the magnetic properties that the doped Ti ions are non-magnetic, namely, they have the $d^0$ electronic configuration \cite{seki}. 
The small intensity discrepancy between the $x$=0.3 and the calculated spectrum at $\sim 461.5$ eV may be due to the splitting of the $e_g$ band, structured effects on the electronic structure as in TiO$_2$ \cite{PRB_04_Kucheyev}, and/or a small amount of Ti$^{3+}$ as in La$_{1-x}$Sr$_x$TiO$_3$ \cite{abbate}.

\begin{figure}[t!]
\begin{center}
\includegraphics[width=8.7cm]{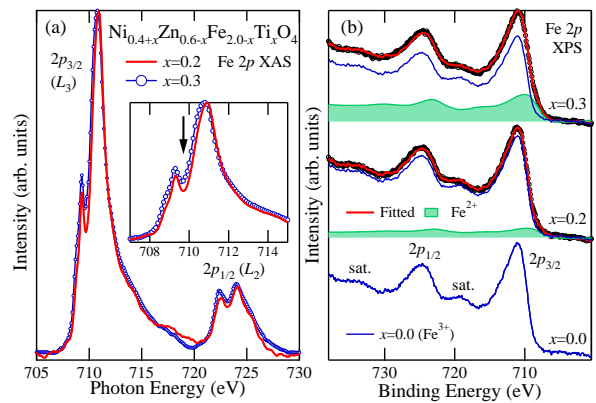}
\caption{Fe $2p$ core-level spectra of NZFT thin films. 
(a) Fe $2p$ XAS spectra. The inset shows an enlarged plot at the $L_3$ edge. The arrow shows a peak position of the Fe$^{2+}$ spectrum. 
(b) Fe $2p$ XPS spectra. The $x$=0.2 and $x$=0.3 spectra have been fitted to a linear combination of the spectrum of $x$=0.0 and that of FeO \cite{graat}. 
}
\label{Fe2p}
\end{center}
\end{figure}

Figure~\ref{Fe2p} shows the Ti-composition dependence of the Fe $2p$ spectra of the NZFT thin films. 
The line shape of the Fe $2p$ XAS spectrum shows a two-peak structure at the $2p_{3/2}$ ($L_{3}$) edge characteristic of trivalent Fe under octahedral crystal field (Fe$^{3+}$ $O_h$) and is similar to those of Fe$^{3+}$ oxides \cite{graat, van}, indicating that most of the Fe ions are trivalent. 
In contrast to the Ni, Zn, and Ti $2p$ spectra, the Fe $2p$ spectra showed a systematic dependence on the Ti concentration. As shown in panel (a), the Fe $2p$ XAS spectrum of the $x$=0.2 sample differs from that of the $x$=0.3 one, e.g., the dip of the two-peak structure became shallower with increasing Ti concentration. 
If the Fe$^{2+}$ ions exist, the Fe$^{2+}$ XAS spectrum is expected to show a peak near the dip \cite{PRB_95_Crovombette}. The existence of the Fe$^{2+}$ ions induced by Ti doping, as has been proposed in the previous report \cite{seki}, can therefore explain the filling of the dip in the Fe $2p$ XAS spectra. 
In order to check whether the Fe$^{2+}$ ions exist or not more clearly, we analyze the changes of the Fe $2p$ XPS spectra with Ti doping, as shown in Fig.~\ref{Fe2p}(b). Without Ti doping, the Fe $2p$ core-level spectrum is close to that of Fe$^{3+}$ oxides \cite{graat}. With increasing Ti concentration, the lower binding energy side of the $2p_{3/2}$ main peak (at $\sim711$ eV) becomes broader and the satellite structure between the $2p_{1/2}$ and $2p_{3/2}$ becomes stronger. 
The $x$=0.2 and $x$=0.3 spectra have been fitted to a linear combination of the spectrum of the $x$=0.0 and that of the FeO (Fe$^{2+}$ $O_h$)~\cite{graat}. 
The fitted spectra well reproduce the Fe $2p$ XPS spectra as shown in Fig.~\ref{Fe2p}(b), indicating the existence of Fe$^{2+}$ ions in the Ti-doped samples. 
The intensity of the Fe$^{2+}$ component increases with increasing Ti concentration, suggesting that the Ti doping into the NZFT thin films induces an overlapping Fe$^{2+}$ component in the Fe $2p$ XAS and XPS spectra. 
The ratio Fe$^{2+}$/Fe$^{3+}$ estimated from the fitting coefficients is $\sim 8 \pm 3$\% and $\sim 23 \pm 5$\% for the $x$=0.2 and the $x$=0.3 sample, respectively, and reasonably agrees well with $\sim 11$\% for $x$=0.2 and $\sim 18$\% for $x$=0.3 expected from the chemical compositions. 
The analysis clearly indicate that the Ti doping converts part of the Fe$^{3+}$ ions into Fe$^{2+}$.

\begin{figure}[t!]
\begin{center}
\includegraphics[width=6.5cm]{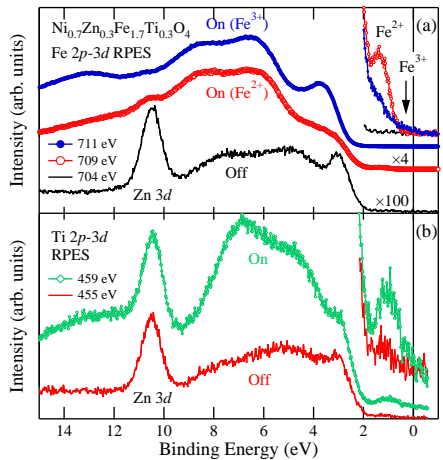}
\caption{Valence-band resonance photoemission spectra of NZFT thin films. 
(a) Fe $2p \to 3d$ resonance photoemission spectra. The photon energies of 709 and 711 eV correspond to the excitation energy for the Fe$^{2+}$ and Fe$^{3+}$ ions, respectively. The off-resonance spectrum was taken with $h\nu = 704$ eV. 
(b) Ti $2p \to 3d$ resonance photoemission spectra. Here, the on- and off-resonance spectra were taken with $h\nu = 459.5$ and 455 eV. In both panels, an enlarged plot near the Fermi level is also shown. 
}
\label{RPES}
\end{center}
\end{figure}

Finally, we shall show that the existence of the Fe$^{2+}$ state is also reflected on the valence-band PES spectra. 
Figure~\ref{RPES} shows resonant photoemission spectra of the $x$=0.3 thin film. 
Based on the XAS and XPS results, there are two kinds of Fe ions, namely, Fe$^{3+}$ and Fe$^{2+}$. Therefore, we have attempted to extract the Fe$^{2+}$ and Fe$^{3+}$ partial density of states (PDOS) separately by tuning the photon energy at $h\nu=709$ and 711 eV for Fe$^{2+}$ and Fe$^{3+}$, respectively. 
In fact, there were differences in the Fe $2p \to 3d$ RPES spectrum between the two photon energies, as shown in Fig.~\ref{RPES}(a). It is important to mention here that the "Fe$^{2+}$ PDOS" show a peak at $\sim1.5$ eV and the "Fe$^{3+}$ PDOS" has finite intensity near the Fermi level. 
This observation suggests that, when electrons are doped into NZFT through Ti doping, the Fe$^{3+}$ states becomes partially itinerant while the number of Fe$^{2+}$ ions increases, consistent with the increase of the Fe$^{2+}$ intensities in the XAS and XPS spectra with Ti concentration as described above. 
In addition, the Ti $2p \to 3d$ RPES spectrum showed a peak at $\sim 1.0$ eV, as shown in panel (b), indicating that the Ti atoms have finite Ti$^{3+}$character probably through hybridization with the Fe$^{2+}$ states or the occupancy of the Ti $3d$ band by a tiny number of electrons. 
The present results are thus consistent with the idea of PIM caused by the inter-valence charge transfer $\mathrm{Ti}^{4+} + \mathrm{Fe}^{2+} \to \mathrm{Ti}^{3+} + \mathrm{Fe}^{3+}$ in NZFT.

In conclusion, we have performed XPS, XAS and RPES studies of (Ni,Zn,Fe,Ti)$_3$O$_4$ thin films with various Ti concentrations. The Ni and Zn $2p$ XPS spectra show that they are divalent and independent of Ti concentration. 
Most of the Ti and Fe ions are tetravalent and trivalent, respectively. 
The XPS and RPES results suggest that the Fe$^{2+}$ ions appear by Ti substitution in (Ni,Zn,Fe)$_3$O$_4$. 
The Fe$^{3+}$ PDOS shows a finite intensity near the Fermi level and differs from the Fe$^{2+}$ one. The Ti $2p \to 3d$ RPES spectrum shows a peak at $\sim 1.0$ eV, suggesting hybridization between the Ti $3d$ and Fe$^{2+}$ states or a slight amount of Ti$^{3+}$, too. 
These results support the proposition by Seki $et$ $ al$. \cite{seki} that  the inter-valence charge transfer Ti$^{4+}$+Fe$^{2+}$ $\rightarrow$ Ti$^{3+}+$Fe$^{3+}$ occurs with light irradiation and enhances PIM in the (Ni,Zn,Fe)$_3$O$_4$ thin film. 

This work was supported by a Grant-in-Aid for Scientific Research in Priority Area "Invention of Anomalous Quantum Materials" (16076208). M.K. and M.T. acknowledge support from the Japan Society for the Promotion of Science for Young Scientists.

\end{document}